\documentstyle[11pt,newpasp,twoside, epsfig]{article}
\def\kms{\ifmmode{\,\hbox{km}\,\hbox{s}^{-1}}\else{$\,$km$\,$s$^{-1}$}\fi}
\def\etal{et~al.}

\def\kmsM{km~s$^{-1}\,$Mpc$^{-1}$}
\def\Nbar{\ifmmode\overline{N}\else$\overline{N}$\fi}

\def\mbar{\ifmmode\overline m\else$\overline m$\fi}
\def\lbar{\ifmmode\overline{L}\else$\overline{L}$\fi}

\def\edcomment#1{\iffalse\marginpar{\raggedright\sl#1\/}\else\relax\fi}
\reversemarginpar

\begin{document}
\title{Testing the Supernova, Cepheid, and Early-type Galaxy Distance Scales}
\author{J.\ P.\ Blakeslee}
\affil{Department of Physics and Astronomy, Johns Hopkins University}

\begin{abstract}
I summarize recent work comparing relative distances measured to 
individual galaxies with independent methods.
The comparisons include: 
ground-based surface brightness fluctuation (SBF)
 and fundamental plane distances to 170 galaxies,
distances predicted from galaxy velocities and the inferred gravity
field, HST SBF
measurements to seven early-type hosts of Type Ia supernovae,
and ties of the Cepheid distance scale to early-type galaxies.
Independent calibrations for some methods provide interesting
constraints on the Cepheid zero point.
\end{abstract}

\vspace{-30pt}
\section{Early-type Galaxy Comparisons: SBF versus FP}

The two most frequently applied early-type galaxy distance indicators
are the fundamental plane (FP, and the related $D_n$-$\sigma$) and
surface brightness fluctuations (SBF) methods.  In a recent study
(Blakeslee \etal\ 2001,\,2002), we used $V$- and $I$-band data from the
ground-based SBF Survey (Tonry \etal\ 2001) to calculate FP photometric
parameters for 170 galaxies with velocity dispersions available in the
homogenized SMAC catalogue (Hudson \etal\ 2001).  To our
knowledge, this is the largest galaxy-by-galaxy comparison of different
standard candle/rod distance methods to date. Fig.\,1a shows the
comparison.~~ 

Overall the distance agreement was good, but several
low-luminosity, S0 galaxies had systematically 
low FP distances, probably due in part to younger ages and lower
mass-to-light ratios, although aperture effects may also contribute.
The SBF distances are tied to the Cepheids via measurements in
spiral bulges, while the FP distances are tied to the Hubble flow
via distant clusters; the Hubble constant that results from this 
comparison is $H_0=68$ \kmsM.  However, we also derived independent
distances for these galaxies based on their velocities and the
gravity field inferred from the redshift-space galaxy density;
the resulting comparison with SBF yields $H_0=74$ (Fig.\,1b),
formally discordant at the 2$\,\sigma$ level with the FP-SBF 
result, but within the range of the systematic uncertainties 
in the various ties.

Another interesting facet of this work relates to the ``fluctuation
number'' $\Nbar\equiv \mbar-m_{\rm tot}$, which measures the galaxy
luminosity in units of the weighted mean stellar luminosity.
\Nbar\ correlates tightly with stellar velocity dispersion;
it also correlates with galaxy color and
is independent of Galactic extinction.  Interestingly,
SBF distances calibrated using the properties of \Nbar,
such as those shown in Fig.\,1b,
can be viewed as a hybrid of SBF and FP distances, and may be 
more accurate than those calibrated from galaxy color alone.
We plan to investigate these issues in more detail.

\begin{figure}[t]
\plottwo{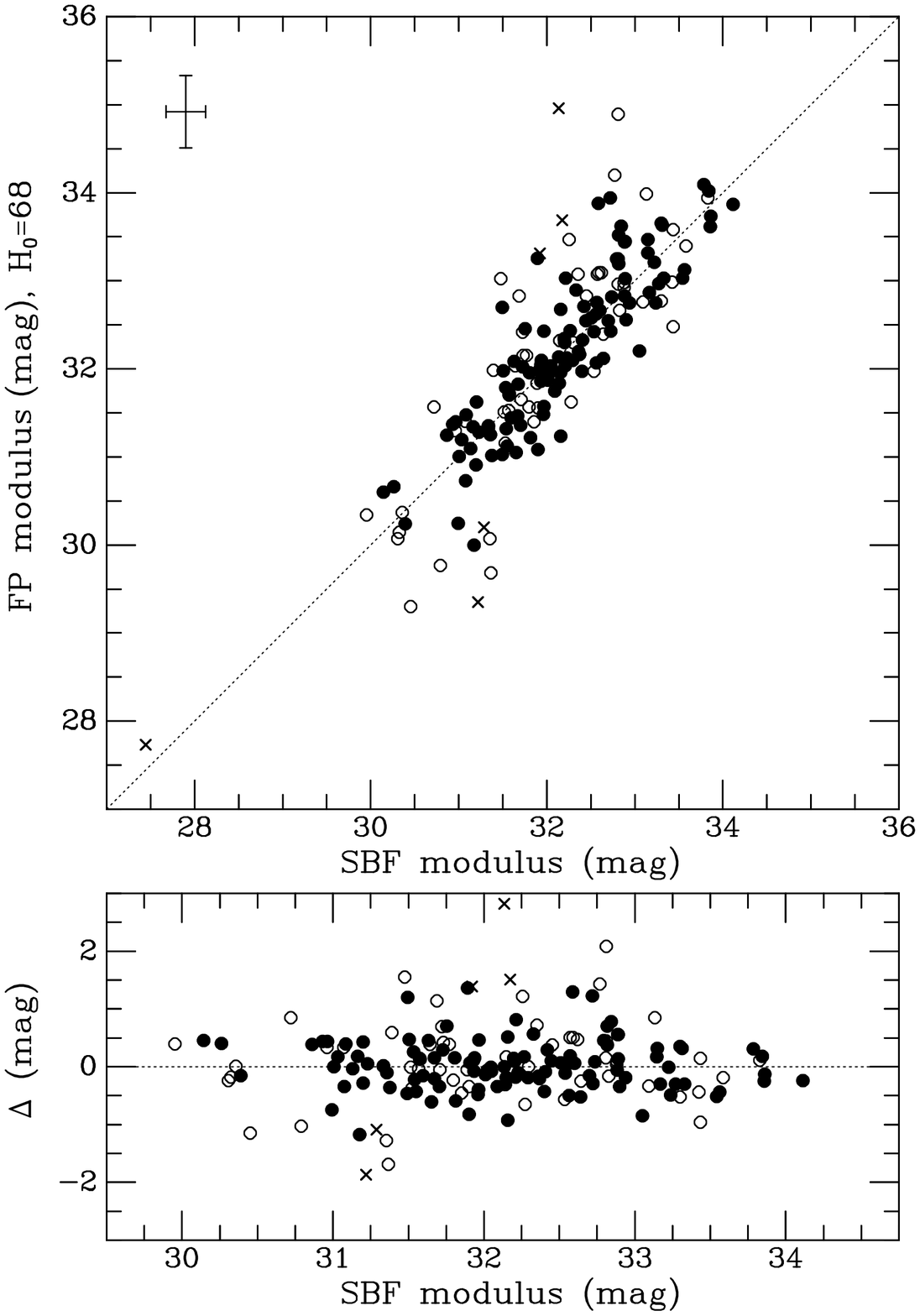}{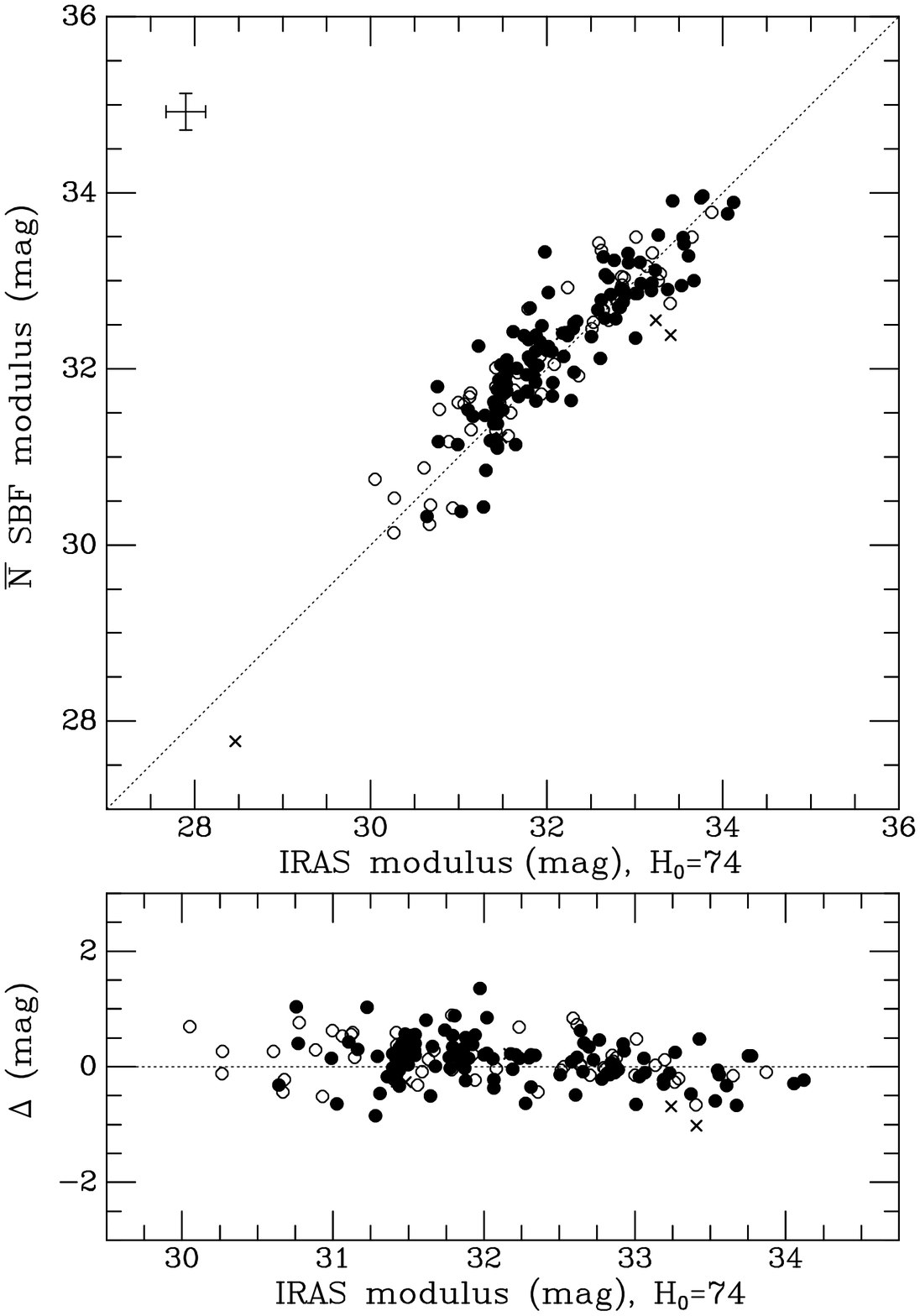}
\caption{\footnotesize{\bf (a)} Comparison of FP and SBF 
distance moduli for the 170 galaxies in the cross-matched 
SBF-SMAC survey samples (from Blakeslee \etal\ 2002). 
The lower panel shows distance residuals.
Filled circles represent true ellipticals, while open circles
represent S0s.  Six galaxies having systematically uncertain
FP or SBF distances are shown as crosses.
{\bf (b)}~Same as (a), but for the comparison of
\Nbar-calibrated SBF distances with those predicted from
the observed galaxy density field in redshift space
(Virgo core galaxies have been assigned the systemic velocity).
\vspace{-7pt}
}
\end{figure}

\vspace{-3pt}
\section{Cepheid Distances to Early-type Galaxies?}
\vspace{-2pt}

Cepheids occur only in spirals and other late-type, star-forming galaxies.
However, the most massive virialized structures in the nearby universe
(e.g., the Virgo, Fornax, and Centaurus clusters), are overwhelmingly
dominated by early-type galaxies.  Although some spirals appear
in projection against the Virgo core, the various secondary indicators
tied to the Cepheid scale indicate that these galaxies are not at
the same distance as the core ellipticals (e.g., Tonry \etal\ 2000;
Ferrarese \etal\ 2000; Kelson \etal\ 2000; Blakeslee \etal\ 2002).
Alternatively, it may be that the secondary indicators are yielding
systematically different results for the calibrating spirals and
the target ellipticals.

We have an ongoing Cycle~10 WFPC2 program to calibrate 
the early-type galaxy distance scale via Cepheid distances
to late-type galaxies that are physically associated with ellipticals.
The target galaxies are the NGC\,4647/NGC\,4649 pair (Fig.\,2a) and
NGC\,5128 (Cen\,A), an elliptical with a central dust lane and 
associated star formation, apparently resulting from the incursion
of a gas-rich dwarf.
The Cen\,A Cepheid observations have yielded more than 60 
superb Cepheids, making this one of largest 
high-quality HST Cepheid data sets.  At present, we are still
finalizing the analysis, but Fig.\,2b shows some example
light curves; differential extinction within Cen\,A is
a major issue for this program.  

\begin{figure}[t]
\plotone{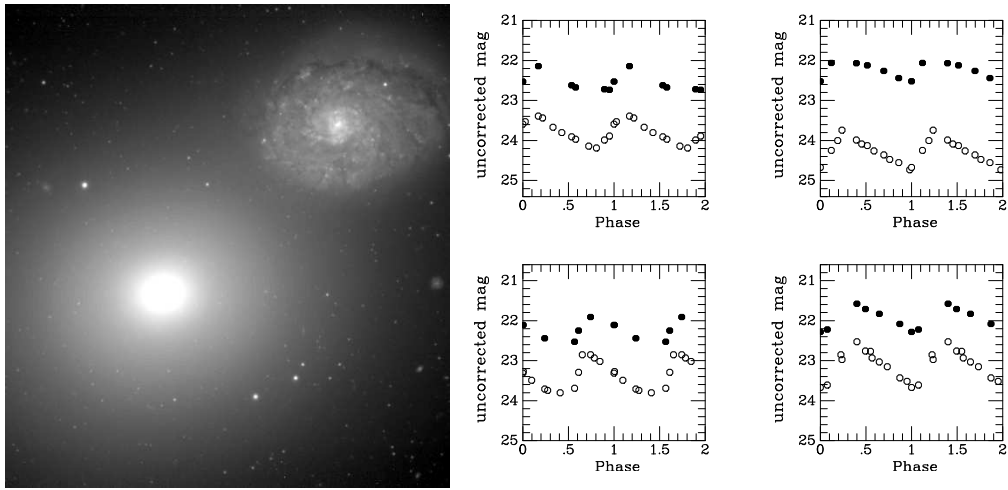}
\caption{\footnotesize{\bf(a)}~NGC\,4649/4647: one of the 
early-/late-type pairs  targeted by our program to calibrate 
elliptical galaxy distances from Cepheids. {\bf(b)}~Some example light
curves for NGC\,5128 (Cen\,A), provided by L.~Ferrarese.
The magnitudes have not been corrected for internal extinction.
\vspace{-7pt}
}
\end{figure}

\vspace{-3pt}
\section{SNe\,Ia versus SBF}
\vspace{-2pt}

SBF and Type Ia supernovae (SNe\,Ia) studies have in the past disagreed on
$H_0$ at the $\sim\,$20\% level, which is surprising for two methods that
routinely achieve 5--10\% internal accuracy.  The excellent resolution of
HST provides an enormous advantage over ground-based data for SBF studies,
and we have recently used WFPC2 to measure high-quality HST/WFPC2 SBF
distances to seven early-type galaxies that have hosted well-observed
SNe\,Ia (Ajhar \etal\ 2001).

The results showed excellent agreement in the relative distances, 
but an offset of $\sim\,$0.25 mag in zero points, which we traced 
to the different, and indeed dissonant,
compilations of Cepheid distances used in the past for the respective
zero-point calibrations of the two methods.  
When calibrated consistently, SBF and SNe\,Ia also
agree in an absolute sense (Fig.\,3a) and give $H_0\approx73$.
This is the first time
the agreement has been demonstrated through a direct comparison of
statistically significant samples of SBF and SNe\,Ia galaxy distances.

\vspace{-3pt}
\section{The Zero Point Problem}
\vspace{-2pt}

The most pressing problem in the measurement of extragalactic
distances appears to be systematic uncertainties in the zero points.
We have seen that there is significant uncertainty in the zero-point
tie of the early-type galaxy distance scale to Cepheids, but perhaps
even greater is the uncertainty in the Cepheid zero point itself, 
in part due to the poorly-constrained LMC distance.

Stellar population models can be used
to predict SBF magnitudes and colors for a large range of metallicities and
ages (e.g., Blakeslee, Vazdekis, \& Ajhar 2001b).  These models reproduce the
observed SBF colors and behaviors very well,
but predict an SBF zero point fainter than the Cepheid-calibrated one by
$0.2{\,\pm\,}0.1\,$mag in $I$ (the only band in which SBF is directly
tied to the Cepheids via spiral bulges).  However, the model and empirical zero
points would come into close agreement if the Cepheid scale were revised
to agree with the dynamical distance to the NGC\,4258 water maser
(Herrnstein \etal\ 1999), for example, by changing the assumed LMC distance
modulus from 18.5 to 18.3 mag. Further refinements of the models
should provide more stringent tests of the distance scale and
guide future SBF programs (Fig.\,3b).

\begin{figure}[t]
\plotone{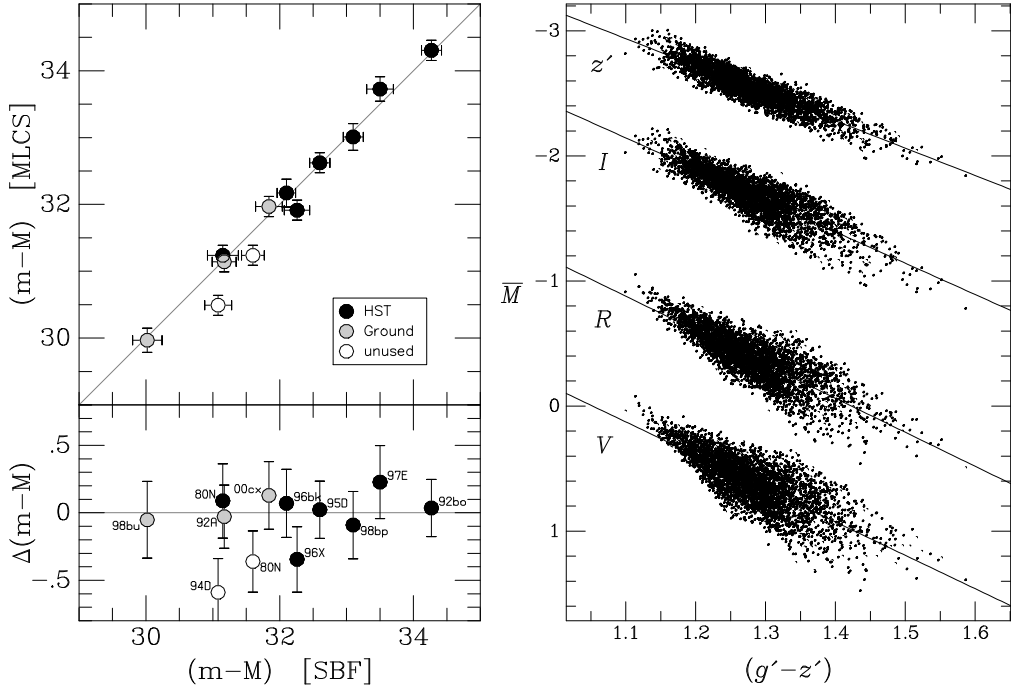}
\caption{\footnotesize
{\bf(a)}~SNe\,Ia distance moduli from the multi-color light curve shape
method are plotted versus SBF distance moduli.
The two ``unused'' galaxies had SNe with
unusual light curves.  
Both methods are tied 
to the ``final''  Key Project Cepheid distances (Freedman \etal\ 2001).
{\bf(b)}~Absolute SBF magnitude $\overline M$ in various bands
is plotted against $(g'{-}z')$ color index for
the composite stellar population models of Blakeslee \etal\ (2001b).
The models indicate that the Cepheid distance scale
should be revised down by 0.1--0.2 mag. Interestingly, they also
predict that $z'$ should be the best optical bandpass for SBF.
\vspace{-8pt}
}
\end{figure}

\acknowledgements
The projects reviewed in this paper have all been group
efforts, and I thank my many distance-scale collaborators.


\vspace{-9pt}
\begin{references}\vspace{-6pt}
{\addtolength{\parskip}{-1pt}
\footnotesize
\reference
Ajhar, E.A., Tonry, J.L., Blakeslee, J.P., Riess, A.G.,
Schmidt, B.P. 2001, \apj, 559, 584
\reference
Blakeslee, J.P., Lucey, J.R., Barris, B.J., Tonry, J.L., 
    Hudson, M.J. 2001a, \mnras, 327, 1004
\reference
Blakeslee, J.P., \etal\ 2002, MNRAS, in press (astro-ph/0111183)
\reference
Blakeslee, J.P., Vazdekis, A., Ajhar, E.A. 2001b, \mnras, 320, 193
\reference
Ferrarese, L., \etal\ 2000, \apj, 529, 745
\reference
Freedman, W.L., \etal\ 2001, \apj, 553, 47
\reference
Herrnstein, J. R., \etal\ 1999, Nature, 400, 539
\reference
Hudson, M.J., Lucey, J.R., Smith, R.J., Schlegel, D.J., Davies, R.L. 2001, \mnras, 327, 265
\reference
Kelson, D.D., et al.\ 2000, \apj, 529, 768
\reference
Tonry, J.L., Blakeslee, J.P.,
Ajhar, E.A., Dressler, A. 2000, \apj, 530, 625
\reference
Tonry, J.L., \etal\ 
2001, \apj, 546, 681

\addtolength{\parskip}{1pt}}
\end{references}
\end{document}